# Creating and troubleshooting microscopy analysis workflows: common challenges and common solutions

Beth A Cimini[1]*

[1]=Broad Institute of MIT and Harvard, Cambridge, MA, USA
* = To whom correspondence should be addressed; Contact details:

Dr. Beth Cimini,
Imaging Platform,
Broad Institute,
415 Main St,
Cambridge, MA 02142
Email: bcimini@broadinstitute.org
Phone: 617-714-7000

# Abstract

As microscopy diversifies and becomes ever-more complex, the problem of quantification of microscopy images has emerged as a major roadblock for many researchers. All researchers must face certainchallenges in turning microscopy images into answers, independent of their scientific question and the images they've generated. Challenges may arise at many stages throughout the analysis process, including handling of the image files, image pre-processing, object finding, or measurement, and statistical analysis. While the exact solution required for each obstacle will be problem-specific, by understanding tools and tradeoffs, optimizing data quality, breaking workflows and data sets into chunks, talking to experts, and thoroughly documenting what has been done, analysts at any experience level can learn to overcome these challenges and create better and easier image analyses.

# Introduction

There are few constants across microscopy's long and varied history, except perhaps for the original goal: to make sense of the world that is smaller than what our eyes can perceive. Early microscopy began as descriptions of the natural world, followed by hypothesis generation; it is therefore no surprise that even well after the invention of the photomultiplier tube and the digital camera, microscopy's stated conclusions have often rested on "representative image shown". In our current quantitative scientific era, microscopists must face the challenge of *image analysis*: turning the astonishing variety of things people do with microscopes into quantitative data. Here, we review several common challenges in image handling and analysis, common themes in approaching difficulties that may arise in carrying out image analyses, and discuss community efforts available to help image analysis learners; while many examples here are pulled from the field of light microscopy of biological samples, general principles apply across disciplines and microscope types.

## Common sources of image analysis challenges

    We review here several common image analysis challenges when carrying out image analysis workflows. Not all experiments will have all sources, but many experiments will have many of these sources, each of which must be addressed for analyses to be properly interpreted. These are summarized graphically in Figure 1. We do not deeply discuss here challenges in sample preparation and/or image acquisition, but of course many nuanced decisions and complexities exist in those steps as well; consideration of those steps is critical in a fully quantitative imaging experiment[1,2], and the measurements one wishes to make should always drive technique selection and optimization at the bench and the microscope. The best solution for some of the difficulties discussed below may be to return to those early stages and create different images, emphasizing the importance of incorporating analysis from the earliest pilot stages of image creation in order to not waste precious samples and perhaps-more-precious human time on creating images that can't answer the scientific question of interest.

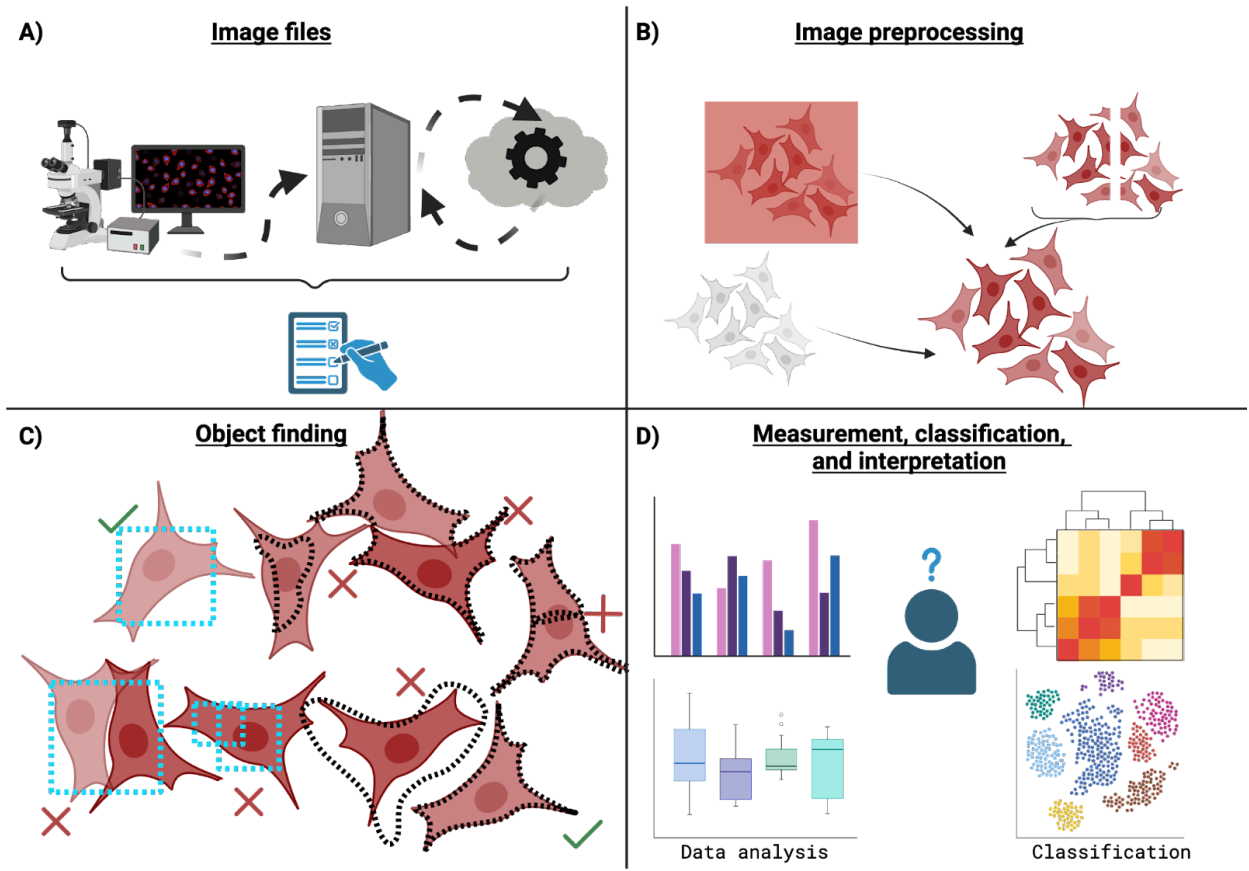

*Figure 1: Common types of image analysis challenges. A: to analyze the image files, analysts may need to navigate proprietary file format export, issues with data size, location where the data must be processed, as well as associating and tracking metadata for future reporting. B: images may need to be pre-processed before analysis, such as by stitching, denoising/background removal, or pixel classification. C: the analyst must determine if they wish to perform object detection to identify the number and approximate location of objects (left side, blue boxes) or instance segmentation to determine exact boundaries (right side, black outlines); see below for further discussion of these terms. In either case, objects may be erroneously merged or split, and segmented objects may not be identified with accurate boundaries. D: the analyst must determine the appropriate ways to normalize, group, and present the data (left). If performing classification, the analyst must decide between supervised and discrete classifications (top right) versus more continuous and/or unsupervised groupings (bottom right).*

## Image files

The first major task a user must usually accomplish in an image analysis workflow is in the handling of the image files themselves. Many microscopes *export* data in proprietary formats that make it difficult to open the files outside of specialist programs; care must also be taken that the export settings do not alter or destroy image data. Photographs (or so-called *natural images*) are typically made up of three images (one red, one green, and one blue) or *channels* with 256 ($2^8$) possible intensity values in each color at each pixel; such images are often called

8-bit (for 2^8) RGB (for the three colors used) images or occasionally 24-bit (3x8) images. Microscope images may contain up to hundreds of channels, and especially across different modalities may capture many more than 256 intensity values (often 4096 (2^12) or 65536 (2^16)) per channel; export software if not set carefully may be optimizing for 8-bit RGB images, which are more easily visualized using standard photograph viewing software but which will then inadvertently cause channel loss in images with more than three channels and clipping and/or compression of intensity values beyond 256. Export software can also sometimes select so-called "lossy" compression, which makes file sizes smaller but can introduce artifactual shapes and/or colors into the data; tiff is a typically safe default choice, though read your software's manual carefully to understand how your data will be treated during export with your chosen settings.

Once exported, one must deal with the size and structure of the data - how many files are present, and in what configuration? Some microscopes create large single files; others (such as some formats associated with slide-scanning microscopes) create many files per scan, which must be kept in a certain relationship to one another or file reading will break. The user must determine how much data has been generated, where it will be immediately stored, if sufficient computational power exists on the immediate storage machine to perform required post-processing and analysis, and what (if any) plans exist for long-term data storage. Creation of a data management plan is critical, and as early as possible[3], and one must consider the appropriateness of location, licensing, etc, especially for sensitive data such as patient samples where consent and anonymization are a factor.

A key and underappreciated aspect of image file handling is the handling of associated *metadata* - how was the sample generated, how was it imaged, and with what experimental question in mind. The answers must be carefully tracked, and ideally permanently associated as closely as possible to the image data so that in the future, it is easy to determine what any image depicts and how it was made. This facilitates not only maximally correct analysis at the time, but eventual data reuse [4].

## Image preprocessing

Once the user has their data, it will often (though not always) require some amount of preprocessing before eventual analysis (Figure 1, top right). If the microscopy method used starts with multiple individual images (such as in single-molecule localization microscopy (SMLM), slide-scanning microscopy, multi-objective image capture methods, or highly-multiplexed imaging methods), customized integration may need to be done to create one "logical image" per sample. Highly scattering samples (such as thick tissue sections) may require *deconvolution* to accurately resolve some structures, especially in widefield microscopy; since fluorescence microscopists typically try to minimize the light on their sample to reduce photodamage and bleaching, *denoising* may be needed to enhance features of interest when the pixel intensity in stained regions is not much higher than in unstained regions. The user may entirely lack a stain for particular regions of interest in the image, or decide the existing stain is insufficiently specific; in these cases, *semantic segmentation* (often called "pixel classification") tools may be called on to create machine learning algorithms that will allow the user to teach the software how to find regions of interest that they wish to measure later. A list of available open source tools for these tasks and others have recently been compiled.[5]

## Object finding

Many microscopy analyses rely on identifying objects within the image. Computer scientists differentiate between *object detection* (finding how many objects are present, typically with a centroid and perhaps a bounding box of where the object can be found) and *instance segmentation* (often in the life sciences simply referred to as "segmentation", despite the existence of other forms of segmentation such as semantic and panoptic segmentation) where it is important to find the exact boundary of the object (Figure 1, bottom left). Users first must decide which kind of object finding to perform: workflows which rely primarily on counting and classification ("how many cells here are infected, and how many uninfected?") are suited to object detection; cases where the user wants to know properties of the specific objects ("how big are infected cells and uninfected cells?") require segmentation. Segmentation is typically considered to be a much more difficult problem than object detection, since it requires much more precision, but it is far more commonly used since it can ultimately provide more information.

Segmentation can typically be performed either using standard computer vision techniques or newer *deep learning* approaches involving *neural networks*, which can solve more difficult segmentation tasks (since with sufficient training data they can learn to avoid many image problems such as debris, changes in staining conditions, and changes in lighting conditions) but require an initial training with large numbers of images and substantial computational resources [6]. Once deep learning networks are trained, they typically do not require substantial compute to run on new data (a task often referred to as *inference*), but they may require expertise to run and they may be very specifically tuned a to specific expectation of input data. Recent advancements in both classical and deep-learning have been recently reviewed[5,7], and efforts are being made in a number of communities to provide tools[8] and so-called model-zoos to improve accessibility of deep-learning approaches to non-computational users[9].

Methods based in classical computer vision typically require that the objects of interest are bright and everything else in the image is dark, which if not already the case requires image pre-processing steps (see above). If this pre-processing is onerous, the user can consider using deep learning techniques for either object detection or segmentation, but in the absence of an existing pre-trained model for the user's task (or the ability to easily create a fine-tuned model, such as with iterative training or retraining)[10,11], training such a network can require substantial data, a considerable time investment in creating the annotated images needed to train the model, and computational expertise.

## Measurement, classification, and interpretation

Once all image preprocessing steps are complete and any objects of interest have been found, the user is finally ready to use their images to answer questions (Figure 1, bottom right). Major sources of complexity at this step begin with simply deciding on the exact metrics to use - does one want to know the total amount of stain in an image or object? The mean amount? How the distribution of the stain has changed? Determining what each metric precisely means, and which is the best match to the scientific question of interest, can take significant knowledge and/or expertise. Statistical treatment of the data also requires a careful approach - is the appropriate unit of comparison an object, an image, a replicate, an organism, a capture date, or some other unit? Does the data need to be normalized for cross-batch comparison, and if so, what constitutes a batch and how will the normalization be performed?

If performing classification analyses, the user must consider if their phenomenon is best represented by a discrete measurement, as is appropriate when there are a finite number of distinct separable phenotypes,  or a continuous measurement, such as when there is smooth progression between states along a trajectory. Different classification methods or techniques may be more appropriate to each class of problem. Guidance as to how to create and interpret various kinds of measurments is available[12].

# Common themes for approaching image analysis challenges

While deep-dive examinations of how to approach all of the possible difficulties described above are beyond the scope of this work, several general principles about *how to solve for any given challenge* can be derived.

## Keep analysis in mind from the very beginning

Whether due to the historical baggage of "representative image shown", personal computational discomfort, or simply lack of habit, many researchers simply don't begin to think about analysis until all the data has been collected. As will be laid out below, understanding the scientific question of interest is critical to assessing data quality; one can *hope*, but cannot truly be *confident*, that their data contains the answer to the scientific question until one actually tries to extract it. This is not to say one can never change one's scientific question, or reuse existing data to answer another question; high-quality data with good metadata annotation can be reused in sometimes wildly different ways [13], and few metrics are so unique and sensitive that there is only a single correct way to get the right data to measure them. Ultimately, though, every aspect of the experiment (which samples are generated, which samples are collected and/or processed together for maximum comparibility, how sample processing proceeds, what imaging technique is selected, which microscope parameters are used) impacts which conclusions can be reliably drawn when the final analysis is done; thus, keeping the analysis in mind during all those stages protects against accidental self-sabotage of one's future ability to fully use the data one worked so hard to make.

## Optimize data quality

It may sound obvious, but a general critical factor in image analysis and computing in general is "garbage-in, garbage-out": to generate high-quality analyses, one needs to have high-quality images. Often, visual assessment is sufficient for the very preliminary stages - is the image in focus, not saturated, and reasonably clear of debris? (Figure 2) Are any objects the researcher wishes to computationally identify visible, and are their boundaries defined enough that the researcher can assess if segmentation is proceeding accurately? - but quantitative analysis of pilot data is always recommended before collecting final data. Perfect images are of course impossible to generate (and ultimately not required), but the images should be of good enough quality that it does not push the analysis workflow outside the error tolerances of the problem (see below). As with tool selection, the eventual measurement to be made and its tolerance will guide what good-enough means, further emphasizing the importance of thinking about analysis early in the process: in a hypothetical experiment with a nuclear marker, cell boundary marker, and marker for some other biology (Marker X) where the goal is to assess how much of Marker X is in the nucleus, a dim/blurry cell boundary marker is likely tolerable, but if the goal is to measure the amount of Marker X present at the cell membrane, it likely is not.

Ultimately, all images (and indeed, all experiments) one creates in the lab are models we are attempting to analyze- after performing some set of experiments on some finite number of samples, we are attempting to create a quantitative picture of what the larger world looks like. Ultimately, as in all models, there will be limitations and inaccuracies - as George Box said, "All models are wrong, but some models are useful". Performing an image analysis is therefore not a matter of performing some perfect series of steps, but rather in creating a model that is most correct/least wrong. Error tolerance is a practical part of any image analysis, with the level of tolerance linked to the expected size of the quantitative change: for example, a 10% error tolerance is perfectly acceptable when looking at a 10-fold change but not when quantifying a 20% change. Pilot experiments are therefore critical to understanding the degree of change one can expect in the final experiment and for creating "ground truth" data (such as hand drawn segmentations or manual sorting of samples into those that do vs do not contain a given artifact) that one can use to measure the error and/or variability present in the analysis workflow and determine whether the desired measurement can in practice be made reliably. If the error is too large even after optimization of all analysis workflow parts, it may be necessary to return to the bench or the microscope (Figure 2).

## Understand the tools and their limitations

In sample preparation and microscopy, it is common to balance the cost, work, and fidelity of individual steps of sample preparation and imaging, or even between imaging modalities. The "balancing act" does not end there; one must also in one's experimental design balance how analysis may need to be adjusted due to issues with image data quality (see above), alongside considerations from each part of the analysis wofkflow (Figure 1). When selecting an analysis tool or approach, a similar set of tradeoffs to those encountered at the bench and the microscope must be weighed: does this approach do what I need? How easy is each tool in this approach to install? How easy is it to use? Will it handle multiple steps, such as both preprocessing AND object finding? How easily can the analysis scale from an initial prototype to many images? How easy is it to inspect each step to make sure it is done correctly? How easily can I document what was done for the future so it can be repeated? Depending on the user's needs, comfort level with individual tools and with scripting/coding in general, there are nearly always many correct approaches to a given problem.

It is also important to understand how the tools and the settings within them may affect the final data. As an example, if the researcher's object of interest is usually around 20 µm in diameter and the analyst sets a hard cutoff during object finding that only objects between 10 and 30 µm in diameter are "real" and should be accepted, perturbations or conditions that create >30 µm diameter or <10 µm diameter objects may not be detected, because these objects will be thrown out due to the cutoff. Such a cutoff may still be good and least-wrong if it throws out relatively few real objects relative to many pieces of debris that would otherwise have thrown off the quantification (see Figure 2), but it means the results should not be interpreted to mean that Perturbation X does not create 35-µm-diameter objects. Likewise, analyses performed on individual 2D slices or 2D projections of 3D volumes must be understood in that context, especially for measuring a structure's size, intensity, and/or colocalization across multiple markers[14,15].

When adopting any new tool, it is important to understand the data types a tool expects (such as fluorescence vs brightfield vs EM, individual object crops vs individual slices vs whole volumes, etc.). In all tools but most especially tools using deep learning, it is critical to understand the conditions under which a tool does and does not work well[16]; this can be more challenging in deep learning because the user is not typically manually setting cutoffs as they might be in a more conventional analysis tool, and networks may be tuned too tightly (or *overfit*[16]) to certain data aspects in ways that produce unexpected results, since it is typically difficult to peer inside the "black box" and tell how a network has learned to make a particular decision[17]. While deep learning undoubtedly solves many problems in microscopy that conventional approaches have not [18], it must be used with especial caution. Whether one should try a deep-learning-based tool vs a conventional tool is not a simple answer for most tasks, and will be based on ease of use, performance metrics, and how many conventional-tool-steps a deep learning tool might replace.

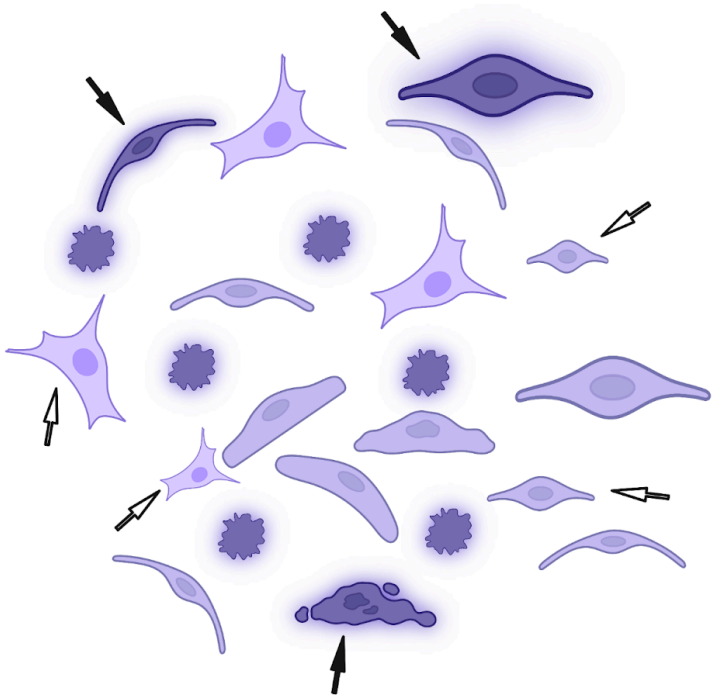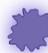

*Figure 2: Example of real-world considerations when removing small, bright debris particles from fluorescent cell images. A few of many possible options are presented; each has both advantages and disadvantages. Color saturation represents brightness of the debris/cell object. Arrows mark cells that would be inadvertently removed from analysis by a debris-removal method due to brightness (filled arrows) or size (unfilled arrows).*

## Work in "right-size", representative chunks

Nearly all image analysis routines are multi-stage workflows: many steps will ultimately take place between the microscope hard drive and a final answer [19]. Especially as data sets get larger and/or the user accumulates more of them, knowing how best to handle complex, compound workflows becomes more and more essential to a high-quality final product and the least-painful experience for the analyst.

Once the basic steps and tool for the workflow have been selected, the workflow should be optimized in pieces, with data quality (see also above) assessed at every stage: in an example workflow that involves alignment, followed by denoising, followed by segmentation, followed by measurement, the aligned images, the denoised images, and the segmentations should all be checked for errors introduced at each step along the way. This optimization typically is best to perform sequentially in the order that the data will travel through the workflow, as changes in later steps will need to be re-assessed anytime an earlier step changes. While it may be tempting to "roll the dice" and only examine the final product to bypass the time spent in quality checks, adding quality control steps along the workflow is ultimately far faster in the long run for the vast majority of cases since it becomes easier to trace the sources of the errors and solve them piecewise.

Unless one's data is extremely small, this prototyping process is made much easier by only working with a subset of the total data set during the prototyping phase. For large images, this may mean a few crops; for sets that consist of many smaller images, a few well-chosen images. It is extremely important that such subset not just consist for example of the first few images captured, the prettiest samples, or the image edges or centers, but instead represents the full range of phenotypes present in the whole data set, or else the analysis workflow may only work on the kind of data in the subset; this is an especially critical factor with deep learning workflows which are prone to *overfitting* to only recognizing certain phenotypes [16]. As a practical example of subsetting, when working in multiwell plates where each well is differently perturbed, pulling one image from each well is often sufficient to ensure one's workflow is robust to the whole experiment before running the final optimized workflow on the full data set.

## Bring together the right experts

It is entirely understandable if a new image analyst has become overwhelmed by this point in the process: it seems like there is far too much for anyone to learn alongside all of the domain-specific and technique-specific knowledge they need to keep up with. This is ultimately true, as image analysis becomes accepted as a discipline in its own right. Understanding these challenges, the open-source image analysis community has created a number of resources in order to help users get started and/or improve their analyses, including tool lists [5] and best practices guides [1,19–21] . In 2018, the Scientific Community Image Forum (forum.image.sc) was launched to create a single central place for users to ask questions related to image analysis [22], and as of mid-2023, serves as a central help forum for >60 individual open-source image analysis tools and contains tens of thousands of posts that are free for users to search for answers. Image analysis also has become an increasingly common option within imaging facilities, and a few stand-alone image analysis facilities now exist [23].

While image analysis experts are critical, image analysis expertise is not the only knowledge needed: expertise in what the samples are, how they were created, and how they were imaged is critical to determining what can and what cannot be learned from any given image. This information may be provided by the researcher, the image analyst, and/or by a microscopy specialist involved in creation of the image. Finally, the most indispensable expert is the researcher, as they are the expert in their scientific question, and therefore which metrics are and are not important to gather and which compromises can be accepted without derailing their analysis. Many local and global organizations now exist to help users figure out how to improve their understanding of imaging and image analysis; a non-exhaustive living list with a focus on bioimaging experiments is available at bioimagingguide.org [1].

## Document everything

Even experts may disagree on the right approach to solving a particular complex problem; image analysis is a constantly-evolving discipline, where new tools and approaches emerge seemingly daily. Ultimately, there is rarely a single "right" answer to any challenge, and certainly there is no single correct workflow for any given class of problem. Ultimately, the correctness of one's analysis rests on the ability of the reader to understand what was done, how, and why. Thorough documentation of every step taken is critical for scientific validation, including metadata of the images, steps taken, programs used, tool versions, order of operations, and beyond. Checklists have been proposed to guide users through the necessary documentation for analyses [20] as well as for the kinds of metadata that are critical to capture in a bioimaging experiment [4], but in general, one will rarely regret the time taken to document an analysis, if only for one's own future understanding when publication time rolls around. Thorough documentation is also critical to reusability and replicability of one's work[24,25].

# Conclusion

It takes many hours of study and work to become an expert in all aspects of image analysis, and in such an ever-changing discipline, by the time one has become an expert, many aspects of one's knowledge have become out-of-date. In such a dynamic field, knowing *how to solve* problems is therefore far more valuable than *knowing all possible solutions*. While it is tempting to think professional image analysts represent a total solution to the problem of the complexity of image analysis, the scientist creating the samples and images must understand what they must contain (and must not contain) in order to do the analysis correctly, and only someone who deeply understands the scientific question being asked can properly balance the various tradeoffs and caveats that inevitably must be weighed in the creation of the analysis workflow. This means that while image analysis professionals are critical in modern science [23], "analysis fluency" from the researchers creating the images is indispensable. Only when the image creators and analysts are in sync (either because they are the same person or because two or more people have learned to communicate well) can one be sure the right answer to the scientific question was reached.

By understanding the steps of analysis (Figure 1), potential pitfalls (Figure 2), working in stages, ensuring high-quality data at every step, understanding how individual steps are shaping the analysis, working with experts, and documenting what was done, even a novice analyst can create high-quality analyses that fairly model the scientific question that their images sought to probe. While exact tools to perform these steps will no doubt change year by year, approaches to solving problems are likely to serve users in good stead for many years to come, and help is available in local and online global resources for those who wish to improve their skills.

# Keywords

image analysis, metadata, deep learning, image processing, object detection, segmentation, workflows, best practices

# Practitioner points

- Image analysis involves a number of possible challenges that may arise between the acquired image and the final answer, including dealing with image files, preprocessing images, finding objects, and measuring and/or classifying the images.
- Several common principles apply to overcoming most kinds of challenges, including to work in stages or "chunks", understanding the tools being used, assessing data quality at every stage of the process, talking to the right experts, and documenting all steps taken.

# Acknowledgements

The author gratefully acknowledges members of the Cimini lab for feedback on this review. Figures were created with BioRender.com.

# Funding

The work was supported by National Institute of General Medical Sciences P41 GM135019 and grant number 2020-225720 from the Chan Zuckerberg Initiative DAF, an advised fund of Silicon Valley Community Foundation. The funders had no role in study design, data collection and analysis, decision to publish, or preparation of the manuscript.

# Competing interest

The author declares that there are no competing interests associated with the manuscript.